\documentclass{article}
\usepackage{spconf}
\usepackage{epsfig}
\usepackage{psfrag}
\usepackage{color}
\usepackage{tabularx}
\usepackage{pstricks, pst-node, pst-plot, pst-circ}
\usepackage{moredefs}

\newcolumntype{C}[1]{>{\centering\arraybackslash}p{#1}}

\def\e{\mathrm{ e}}
\def\PSNR{\mathrm{ PSNR}}
\def\dB{\,\mathrm{ dB}}

\title{Spatio-Temporal Error Concealment in Video by Denoised Temporal Extrapolation Refinement}
\name{J\"urgen~Seiler$^1$, Michael Sch\"oberl$^2$, and Andr\'e~Kaup$^1$}

\address{
\begin{minipage}{0.52\textwidth}\begin{center} $^1$ Multimedia Communications and Signal Processing \\University of Erlangen-Nuremberg\\ Cauerstr. 7, 91058 Erlangen, Germany\end{center}\end{minipage}
\begin{minipage}{0.43\textwidth}\begin{center}$^2$ Fraunhofer IIS\\ Am Wolfsmantel 33\\ 91058 Erlangen, Germany \end{center}\end{minipage}
}

\ninept
\begin{document}
\topmargin=0mm
\maketitle


\begin{abstract} \label{abstract}
In video communication, the concealment of distortions caused by transmission errors is important for allowing for a pleasant visual quality and for reducing error propagation. In this article, Denoised Temporal Extrapolation Refinement is introduced as a novel spatio-temporal error concealment algorithm. The algorithm operates in two steps. First, temporal error concealment is used for obtaining an initial estimate. Afterwards, a spatial denoising algorithm is used for reducing the imperfectness of the temporal extrapolation. For this, Non-Local Means denoising is used which is extended by a spiral scan processing order and is improved by an adaptation step for taking the preliminary temporal extrapolation into account. In doing so, a spatio-temporal error concealment results. By making use of the refinement, a visually noticeable average gain of 1 dB over pure temporal error concealment is possible. With this, the algorithm also is able to clearly outperform other spatio-temporal error concealment algorithms.

\end{abstract}


\begin{keywords}
Error Concealment, Signal Extrapolation, Video Communication
\end{keywords}


\section{Introduction} \label{sec:introduction} 

The amount of video data that is produced every day has steadily increased in the past years. A reason for this is that storage and transportation of video data has become feasible by the use of state-of-the-art video codecs whereas uncompressed transportation of video data is still a challenging task today. However, besides the quality loss introduced by the coding, the compressed data is more prone to errors than an uncompressed format. Thus, if the compressed video data is transmitted over wireless or packet-based channels where the risk of transmission errors is ubiquitous, severe distortions might occur. In order to cope with this, two strategies are used as outlined in \cite{Stockhammer2005}: error resilience on the one hand and error concealment on the other. In this context, error resilience is applied at the encoder and aims at making the coded video more robust against transmission errors at the expense of a decreased coding efficiency. In contrast to this, error concealment is applied at the decoder for reducing the distortion introduced by occurring transmission errors. This is important for two reasons: Firstly, to provide a decent quality despite the transmission error and secondly, to reduce the effect of error propagation which is caused by the predictive structure of video codecs. Since error concealment is only applied at the decoder, it is not standardized and every decoder can use error concealment algorithms according to the available computing power and quality demands.

In literature, a large amount of error concealment algorithms exists. All these algorithms can be divided into three different groups. First, there is spatial error concealment like the simple Maximally Smooth Recovery \cite{Wang1993} or more sophisticated algorithms like one based on Stochastic Factor Graphs (SFG) \cite{Li2008}. All algorithms from this group have in common that they only rely on the available information of the actually distorted frame. The second group only uses information from preceding or succeeding frames and carries out the concealment by locating areas in these frames for replacing the distortions. One algorithm of this group is the widely used Decoder Motion Vector Estimation (DMVE) \cite{Zhang2000a}. The third group of algorithms like the Improved Fading Scheme (IFS) \cite{Hwang2008} or Motion-Compensated Frequency Selective Extrapolation \cite{Seiler2011} exploit temporal as well as spatial information for concealing the transmission errors. Since typical video sequences exhibit strong temporal correlations, the reconstruction quality usually increases from spatial to temporal to spatio-temporal error concealment as shown in \cite{Tsekeridou2000}. However, for exploiting the available correlations effectively, the complexity typically increases accordingly.

In order to keep the complexity of spatio-temporal error concealment manageable, we proposed the Adaptive Joint Spatio-Temporal Error Concealment (AJSTEC) \cite{Seiler2008b} for successively combining spatial and temporal information by first carrying out a temporal extrapolation via DMVE and after that performing spatial refinement by Frequency Selective Extrapolation (FSE) \cite{Seiler2008}. By splitting the spatio-temporal error concealment into these two sub-tasks, a high reconstruction quality was possible. In \cite{Seiler2010b} we further showed that for the prediction step in video coding the concept of spatial refinement is not limited to the use of FSE, but rather can be regarded as a denoising problem. In this paper we show that denoising techniques can also be used for achieving a spatio-temporal error concealment in video and introduce the concept of Denoised Temporal Extrapolation Refinement in the following.


\section{Spatial Refinement by Non-Local Means} \label{sec:nlm-refinement}

The subsequently proposed Denoised Temporal Extrapolation Refinement (DTER) can be regarded as a generic algorithm for spatio-temporal error concealment. For this, the concealment process is split into two parts. First, a temporal extrapolation is conducted and any algorithm belonging to this group can be used for this. The output of the temporal error concealment acts as a preliminary estimate for the desired signal. In the second step, this preliminary estimate is spatially refined by making use of denoising techniques. In doing so, spatial information can be incorporated and one finally gets a spatio-temporal error concealment.

\begin{figure}
	\psfrag{m}[t][t][0.9]{$m$}%
	\psfrag{n}[t][t][0.9]{$n$}%
	\psfrag{x}[t][t][0.9]{$x$}%
	\psfrag{y}[t][t][0.9]{$y$}%
	\psfrag{t}[t][t][0.9]{$t$}%
	\psfrag{B}[l][l][0.9]{$\mathcal{B}$}%
	\psfrag{A}[l][l][0.9]{$\mathcal{A}$}%
	\psfrag{x0}[t][t][0.9]{$x_0$}%
	\psfrag{y0}[t][t][0.9]{$y_0$}%
	\psfrag{tau}[t][t][0.9]{$\tau$}%
	\psfrag{tau1}[t][t][0.9]{$\tau-1$}%
	\centering
	\includegraphics[scale=0.275]{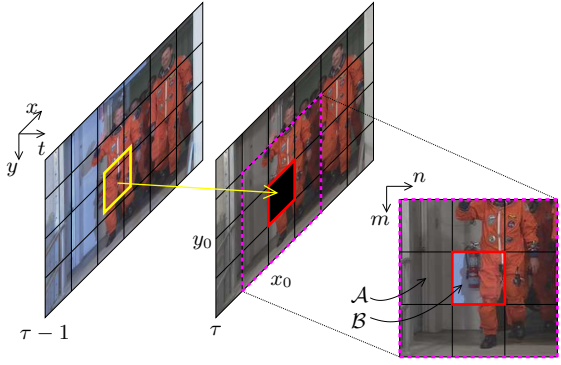}\vspace{-2mm}
	\caption{Relationship between distorted video sequence and processing area $\mathcal{L}$ for concealing the block located at $\left(x_0, y_0\right)$. $\mathcal{L}$ consists of known samples in area $\mathcal{A}$ and temporally extrapolated samples \mbox{in $\mathcal{B}$}.}\vspace{-2mm}
	\label{fig:refinement_area}
\end{figure}

In order to outline DTER, the video sequence $v\left[x,y,t\right]$ with spatial coordinates $x,y$ and temporal coordinate $t$ is regarded. It is now assumed that a transmission error has happened, resulting in a lost block at position $\left(x_0,y_0\right)$ in frame $t=\tau$. Fig.\ \ref{fig:refinement_area} shows the impaired frame and its preceding frame as an example. The lost block now has to be concealed. The consideration of a single isolated block loss is only for presentational reasons. In general, the distorted area can be of arbitrary shape. As mentioned above, first a temporal extrapolation is carried out, that is to say, in a preceding frame a region is determined for replacing the unknown area in frame $t=\tau$. For this, in general any temporal error concealment algorithm can be used. However, in the following, DMVE \cite{Zhang2000a} will be applied for this. Unfortunately, the signal within the block from the preceding frame may not fit to the neighboring correctly received blocks as shown in the example in Fig.\ \ref{fig:refinement_area}. In order to allay this problem, spatial refinement is introduced. For this, the correctly received samples around the lost block are regarded. Together with the preliminary temporal estimate, they form the processing area $\mathcal{L}$. This area consists of correctly received samples from frame $t=\tau$, subsumed in area $\mathcal{A}$, and the preliminary temporal estimate from frame $t=\tau-1$, subsumed in area $\mathcal{B}$. The signal in area $\mathcal{L}$ is described by $s\left[m,n\right]$ with coordinates $m$ and $n$. The relationship between video sequence $v\left[x,y,t\right]$ and processing area $\mathcal{L}$ is also shown in Fig.\ \ref{fig:refinement_area}.

With $o\left[m,n\right]$ being the desired but unavailable original signal, the available signal $s\left[m,n\right]$ in area $\mathcal{L}$ can be regarded as
\begin{equation}
	s\left[m,n\right] = \left\{ \begin{array}{ll} o\left[m,n\right] & \forall \left(m,n\right)\in \mathcal{A} \\ o\left[m,n\right] + t_\mathrm{e}\left[m,n\right] & \forall \left(m,n\right)\in \mathcal{B} \end{array}\right. .
\end{equation}
In this context, $t_\mathrm{e}\left[m,n\right]$ can be regarded as an error term introduced by the imperfectness of the temporal extrapolation. Hence, for improving the quality of the concealment, $t_\mathrm{e}\left[m,n\right]$ should be reduced as much as possible. To achieve this, $t_\mathrm{e}\left[m,n\right]$ is regarded as noise term which should be removed. Consequently, spatial refinement can be regarded as denoising problem and in general any denoising algorithm can be used for improving the error concealment performance. In the following, Non-Local Means (NLM) denoising \cite{Buades2005} is considered. This algorithm is well suited since it effectively exploits self similarities within a signal. Due to this, the spatial correlations within a video signal can be taken into account on top of the temporal correlations which have been exploited by DMVE.

\begin{figure}
	\centering
	\includegraphics[scale=0.275]{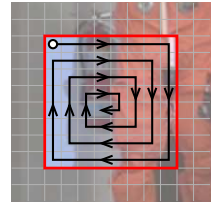}\vspace{-2mm}
	\caption{Example of spiral scan order for processing samples to be refined in area $\mathcal{B}$ (marked in red).}\vspace{-2mm}
	\label{fig:spiral_scan}
\end{figure}

For carrying out the refinement, NLM is performed on the union of the undistorted samples in $\mathcal{A}$ and the temporally extrapolated samples in $\mathcal{B}$. As only samples in region $\mathcal{B}$ are impaired, the refined signal $\hat{s}\left[m,n\right]$ has to be determined only in this area. The processing of the samples in area $\mathcal{B}$ is carried out sample-wise. In this process, samples that already have been refined can be used as reference for refining succeeding ones. Hence, the actual order for processing the samples is important. One possibility would be to process the samples in line scan order. However, in order to more effectively draw spatial information into area $\mathcal{B}$, a spiral scan from the margin of area $\mathcal{B}$ to its center is used as shown exemplary in Fig.\ \ref{fig:spiral_scan}. 

\begin{figure}
	\psfrag{sm}[l][l][0.9]{$s\left[m,n\right]$}%
	\psfrag{sk1}[l][l][0.9]{$s\left[k_1,l_1\right]$}%
	\psfrag{sk2}[l][l][0.9]{$s\left[k_2,l_2\right]$}%
	\centering
	\includegraphics[scale=0.275]{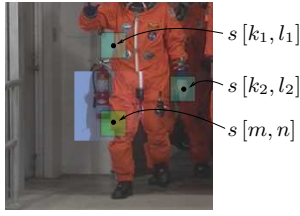}\vspace{-2mm}
	\caption{Example for refinement of pixel $s\left[m,n\right]$. Since the neighborhood of samples like $s\left[k_1,l_1\right]$ is similar, these samples get high weights, while ones like $s\left[k_2,l_2\right]$ with a distinct neighborhood only obtain small weights.}\vspace{-2mm}
	\label{fig:nlm_refinement}
\end{figure}

According to \cite{Buades2005}, the denoised, respectively, the refined samples result from a weighted average of all pixels. So, a refined sample $\hat{s}\left[m,n\right]$ is given by\vspace{-1mm}
\begin{equation}
\label{eq:refinement}
 \hat{s}\left[m,n\right]=\frac{\displaystyle\sum_{\left(k,l\right)\in\mathcal{L}} s\left[k,l\right] w_{\left(m,n\right)}\left[k,l\right]}{\displaystyle\sum_{\left(k,l\right)\in\mathcal{L}} w_{\left(m,n\right)}\left[k,l\right]},\ \forall \left(m,n\right)\in\mathcal{B}.\vspace{-1mm}
\end{equation}
The weight $w_{\left(m,n\right)}\left[k,l\right]$ that is used for every pixel during the averaging depends on its own location $\left(k,l\right)$ and on the location $\left(m,n\right)$ of the sample actually to be refined. The summation in the denominator is required only for normalization. In order to calculate the weight for every sample, the sum of squared differences
\begin{equation}
\label{eq:distance}
 d_{\left(m,n\right)}\hspace{-1mm}\left[k,l\right] \hspace{-0.95mm}=\hspace{-0.95mm} \frac{1}{\left(2d_\mathrm{m}+1\right)^2}\hspace{-3mm}\sum_{{\mu, \nu=}\atop{-d_\mathrm{m},\ldots, d_\mathrm{m}}}\hspace{-5mm} \left(s\left[m\hspace{-1.00mm}+\hspace{-1.00mm}\mu,n\hspace{-1.00mm}+\hspace{-1.00mm}\nu\right]\hspace{-1.00mm}-\hspace{-1.00mm}s\left[k\hspace{-1.00mm}+\hspace{-1.00mm}\mu,l\hspace{-1.00mm}+\hspace{-1.00mm}\nu\right]\right)^2
\end{equation} 
between the neighborhood around the pixel $s\left[m,n\right]$ to be refined and the neighborhood around the candidate pixel $s\left[k,l\right]$ is considered. Both neighborhoods have a width of $2d_\mathrm{m}+1$ samples. Only in the case that a sample is located at the border of $\mathcal{L}$, the regarded neighborhoods are not square and have to be shrunk to a size so that no samples lying outside are referenced. Using $d_{\left(m,n\right)}\left[k,l\right]$, the actual weighting results from
\begin{equation}
 \label{eq:weight}
 w_{\left(m,n\right)}\left[k,l\right] = \e^{-d_{\left(m,n\right)}\left[k,l\right] / h^2}.
\end{equation}
Hence, the weighting depends on the similarity between the neighborhood around the pixel to be refined and the regarded candidate pixel. If the neighborhoods are similar, the candidate pixel gets a high weight and therewith strong influence while for a distinct neighborhood, the weighting becomes very small. To give an example for the refinement, Fig.\ \ref{fig:nlm_refinement} shows the weight calculation for one pixel $s\left[m,n\right]$. A sample like $s\left[k_1,l_1\right]$ gets a high weight during the averaging, since its neighborhood is similar to the neighborhood around $s\left[m,n\right]$. In contrast, samples like $s\left[k_2,l_2\right]$ only get a small weight, as their surrounding neighborhood differs strongly from the neighborhood around $s\left[m,n\right]$. However, the weighting does not only depend on $d_{\left(m,n\right)}\left[k,l\right]$, but further includes the term $h^2$.

The parameter $h$ controls the strength of the averaging and therewith of the spatial refinement. For large values of $h$, a sample gets a strong weighting even if its surrounding neighborhood is very distinct from the neighborhood of the pixel to be refined. If $h$ becomes small, only samples with a very similar neighborhood get a significant weighting. One possibility would be to select a fixed value of $h$ for the refinement. However, this strategy is not good since the strength of the refinement should depend on the quality of the preliminary temporal estimate. In order to achieve this, we propose to calculate $h$ adaptively with a good temporal estimation resulting in a small $h$ and a poor temporal estimation resulting in a large $h$.

\begin{figure}
	\psfrag{m}[t][t][0.9]{$m$}%
	\psfrag{n}[t][t][0.9]{$n$}%
	\psfrag{x}[t][t][0.9]{$x$}%
	\psfrag{y}[t][t][0.9]{$y$}%
	\psfrag{t}[t][t][0.9]{$t$}%
	\psfrag{D}[l][l][0.9]{\color{green}$\mathcal{D}$}%
	\psfrag{x0}[t][t][0.9]{$x_0$}%
	\psfrag{y0}[t][t][0.9]{$y_0$}%
	\psfrag{x0vx}[t][t][0.9]{$x_0+v_\mathrm{x}$}%
	\psfrag{y0vy}[t][t][0.9]{$y_0+v_\mathrm{y}$}%
	\psfrag{tau}[t][t][0.9]{$\tau$}%
	\psfrag{tau1}[t][t][0.9]{$\tau-1$}%
	\centering \vspace{-2mm}
	\includegraphics[scale=0.275]{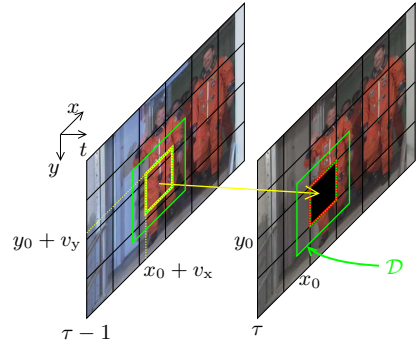}\vspace{-2mm}
	\caption{Test area $\mathcal{D}$ for evaluating quality of temporal extrapolation of an unknown block at location $\left(x_0,y_0\right)$. \vspace{-4mm}}
	\label{fig:decision_area}
\end{figure}

For quantifying the quality of the preliminary temporal extrapolation, a so-called test area $\mathcal{D}$ around the lost block is regarded. This area contains only undistorted samples and is shown exemplary in Fig.\ \ref{fig:decision_area}.  With $v_\mathrm{x}$ being the relative horizontal displacement of the block that serves for the preliminary temporal estimate, and $v_\mathrm{y}$ being the vertical one, the corresponding area can be taken from the reference frame and the temporal error
\begin{equation}
e_\mathcal{D}\left[x_0,y_0\right] \hspace{-1mm} = \hspace{-2.5mm} \sqrt{\hspace{-0.5mm}\frac{1}{\left|\mathcal{D}\right|}\hspace{-1.5mm} \sum_{\left(x,y\right)\in\mathcal{D}}\hspace{-3mm} \left( v\left[x,y,\tau\right] \hspace{-0.5mm}-\hspace{-0.5mm} v\left[x\hspace{-0.5mm}-\hspace{-0.5mm}v_\mathrm{x},y\hspace{-0.5mm}-\hspace{-0.5mm}v_\mathrm{y},\tau-1\right] \right)^2}
\end{equation}
can be calculated for the lost block at position $\left(x_0,y_0\right)$. This metric determines the mean error for the samples in $\mathcal{D}$ if the same displacement is used as for the preliminary estimate of the lost block. In doing so, it indicates how good the temporal extrapolation might match the original signal.

Based on $e_\mathcal{D}\left[x_0,y_0\right]$, parameter $h$ has to be estimated. In this context, a large temporal error should yield a strong averaging, i.\ e., $h$ should become large, whereas for a small error the opposite should become true. For this, we propose the simple relationship 
\begin{equation}
h=\left\{ \begin{array}{ll} e_\mathcal{D}\left[x_0,y_0\right] - \eta & \mbox{if } e_\mathcal{D} > \eta \\ 0 & \mbox{else}\end{array} \right.
\end{equation}
which realizes a linear mapping with a small offset $\eta$ and a limiting of $h$ to positive values. Of course, more sophisticated relationships may be used as well. However, this simple relationship already achieves very good results and is beneficial since it is controlled only by a single parameter. As $h$ depends on the quality of the temporal extrapolation, it has to be calculated only once for every block.

After the refinement of a sample has been finished, it is also inserted back into $s\left[m,n\right]$ for serving as reference for later processed samples. Finally, after all samples in $\mathcal{B}$ have been refined, the refined signal $\hat{s}\left[m,n\right]$ is used for concealing the distortion. This signal then contains spatial as well as temporal information and therewith is superior to a pure temporal error concealment.

\section{Simulation Setup and Results} \label{sec:results}

\begin{figure*}
	\centering 
	\includegraphics[width=0.81\textwidth]{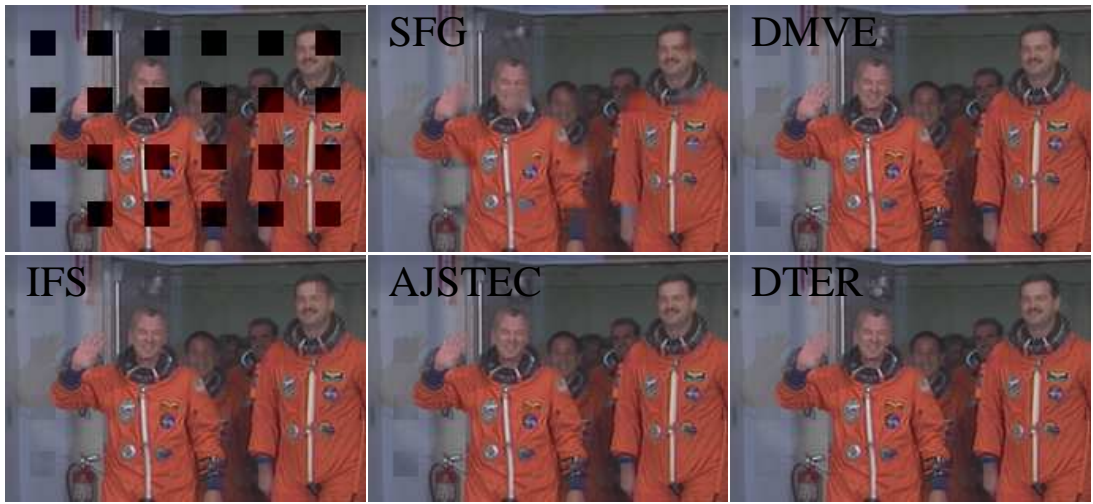} \vspace{-2mm}
	\caption{Visual results for error concealment by the regarded algorithms in a zoomed region of frame $61$ of sequence Crew. The top left image shows a part of the examined error pattern.\vspace{-2mm}}
	\label{fig:visual_results}
\end{figure*}

For evaluating the performance of DTER, the concealment of losses in five CIF sequences is tested. Due to the very generic concept of DTER, it can be used for the concealment of arbitrarily shaped regions in general. However, for the outlined experiments, only isolated block losses of size $16\times 16$ samples in the luminance component are considered in order to allow for a fair comparison to the other algorithms. This error pattern relates to distortions that occur if the Flexible Macroblock Ordering of H.264/AVC is used in combination with the checker board pattern. An example of the loss pattern is shown in the top left of Fig.\ \ref{fig:visual_results}. Four different algorithms are regarded for comparison. This is SFG \cite{Li2008} as a spatial algorithm, DMVE \cite{Zhang2000a} as a temporal one, and IFS \cite{Hwang2008} as representative for spatio-temporal error concealment. In addition to this, AJSTEC \cite{Seiler2008b} is considered, since it also follows the idea of spatial refinement for obtaining a spatio-temporal error concealment.

The search range for all algorithms which rely on some kind of motion estimation is $16$ samples in every direction. For carrying out the refinement, DTER takes a square region $\mathcal{A}$ of $12$ samples width around the lost block into account. Thus, $\mathcal{L}$ is of size $40\times 40$ samples. The width $d_\mathrm{m}$ of the neighborhood around each pixel is set to $6$ pixels, area $\mathcal{D}$ is $8$ samples wide, and the threshold $\eta$ is set to $5$. Fortunately, none of these parameters is very sensitive and all can be varied widely without heavily affecting the reconstruction quality. As outlined in the previous section, the refinement is conducted in spiral scan order and the starting point is the top left corner pixel.

\begin{table}
\caption{Error concealment quality in $\dB$ $\PSNR$ for different sequences and mean gain over DMVE.}\small
\label{tab:psnr_results}
 \setlength{\tabcolsep}{0.7mm}
\begin{tabular}{|c|c|c|c|c|c||c|}
\hline  &  City & Crew & Flower & Foreman & Vimto & Mean gain \\ 
\hline SFG \cite{Li2008} & $22.41$ & $27.30$ & $16.47$  & $23.77$ & $25.31$ & $-6.09$ \\ 
\hline DMVE \cite{Zhang2000a} & $29.52$ & $30.45$ & $26.68$  & $32.74$ & $26.34$ & --- \\ 
\hline IFS \cite{Hwang2008} & $29.29$ & $31.92$ & $23.12$  & $30.50$ & $28.72$ & $-0.44$ \\ 
\hline AJSTEC \cite{Seiler2008b} & $29.36$ & $ 32.64$ & $25.26$  & $ 33.31$ & $28.67$ & $0.70$ \\ 
\hline DTER & $ 29.57$ & $32.23$ & $ 26.76$  & $33.10$ & $ 29.02$ & $0.99$ \\ 
\hline 
\end{tabular}  \vspace{-5mm}
\end{table}

In Table \ref{tab:psnr_results}, the results are listed for concealing the presented error pattern in the first $148$ frames of the test sequences. In this process, the errors firstly occur in the second frame for providing a temporal reference frame for the temporal and spatio-temporal algorithms. The table shows the $\PSNR$ evaluated for the whole sequences and the highest available quality is written bold. Furthermore, the average gain over DMVE is listed for compacting the results. Apparently, the spatial algorithm SFG yields the lowest $\PSNR$ since it does not make use of the temporal correlations within a sequence. Although the spatio-temporal IFS is able to outperform the pure temporal DMVE in some cases, for the regarded test set, it performs worse on average. The table also shows, that with AJSTEC and the proposed DTER, both algorithms that use spatial refinement yield a very high quality and are able to outperform the other algorithms considerably. It can also be discovered that the novel DTER outreaches AJSTEC by an additional $0.3 \dB$ on average, yielding the highest quality among the tested algorithms.

The high quality of DTER can also be seen visually. In \mbox{Fig.\ \ref{fig:visual_results}}, the output of the different algorithms is shown for frame $61$ of sequence Crew. This frame is difficult to conceal since a flashlight is occurring that changes the illumination of the whole scene from one frame to the other. Thus, all algorithms that rely on temporal information have to deal with the problem that the preceding frame is too dark. The pure spatial SFG does not suffer from this problem but it is not able to recover fine details. As DMVE completely relies on the temporal information, the temporally inserted blocks become quite apparent. The spatio-temporal algorithms allay this problem and show less artifacts. Regarding the proposed DTER, it can be seen that it is not able to completely avoid artifacts in this very challenging scenario but among the algorithms it again performs best and it is able to provide a good trade-off between introducing artifacts from the changing illumination and recovering even fine details by exploiting the temporal information.


\section{Conclusion} \label{sec:conclusion}

In this contribution, Denoised Temporal Extrapolation Refinement was introduced for error concealment in video communication. By combining the two widely used techniques of temporal error concealment and spatial denoising, an efficient spatio-temporal concealment of transmission errors is possible, leading to an average gain of $1 \dB$ over a pure temporal concealment. The application of denoising for spatial refinement can be regarded as a generic concept and other denoising algorithms besides NLM can be used for this, as well. Hence, future work will aim at evaluating other denoising algorithms for this purpose and at modeling the temporal extrapolation error better for incorporating this information into the denoising step and therewith into the spatial refinement.



{

}

\end{document}